\documentstyle[aps,epsf,preprint]{revtex}
\newcommand{\fslash}[1]{\ooalign{\hfil/\hfil\crcr$#1$}}

\begin{document}
\draft
\tighten
\title{QCD sum rules for the pseudoscalar decay constants\\
       --- To constrain the strange quark mass }
\author{Hungchong Kim \footnote{E-mail : hung@phya.yonsei.ac.kr}
}
\address{ Institute of Physics and Applied Physics, Yonsei University, 
Seoul 120-749, Korea}

\maketitle
\begin{abstract}

We study the higher order corrections of quark masses to the
Gell-Mann$-$Oakes$-$Renner (GOR) relation
by constructing  
QCD sum rules exclusively for pseudoscalar mesons from
the axial-vector correlation function,
$i \int d^4x~ e^{ip\cdot x} \langle 0| T[A_\mu (x) A_\nu (0)] 
| 0 \rangle $.  To project out the pseudoscalar meson contributions,
we apply $-p^\mu p^\nu/p^2$ to this correlation function and
construct sum rules for 
the decay constants of pseudoscalar mesons, $f_\pi, f_k$ and
$f_{\eta_8}$.  The OPE 
is proportional to quark masses due to PCAC.  
To leading order in quark mass, each sum rule reproduces the corresponding
GOR relation. 
For kaon and $\eta_8$, the deviation from the GOR relation due to
higher orders in quark mass  is found to be substantial.
But the deviation gives better agreements with the
phenomenology.
Our sum rule provides a sensitive relation between $f_K$ and $m_s$,
which stringently constrains the value for $m_s$.
To reproduce the experimental value for $f_K$, $m_s$  
is found to be
186 MeV at 1 GeV scale. The $f_{\eta_8}$ sum rule also supports this finding. 

\end{abstract}

\vspace{30pt}

\section{INTRODUCTION}
\label{sec:intro}

According to PCAC~\cite{adler}, the decay constant of a pion 
measures the coupling
strength of the corresponding axial-vector current to the pion. One
useful constraint which can be derived by combining PCAC with the
soft-pion theorem is the 
Gell-Mann$-$Oakes$-$Renner (GOR) relation
\begin{eqnarray}
-4 m_q \langle {\bar q} q \rangle = m_\pi^2 f_\pi^2\ , 
~(f_\pi =131~{\rm MeV})\ .
\label{gor}
\end{eqnarray}
As the pion decay constant $f_\pi$ is well-known\cite{epjc}, 
this GOR relation can be used to constrain either the quark condensate
or the quark mass. In particular, the small quark mass 
$m_q = (m_u+m_d)/2 \sim 7$ MeV~\cite{gasser,leut} at 1 GeV scale
restricts the quark condensate to be  $\langle {\bar q} q \rangle 
\sim -(225~{\rm MeV})^3$, the value being widely used in practical 
QCD sum rule calculations.
Corrections to the GOR relation are an order of ${\cal O} (m_q^2)$ or 
higher, which should be small.

When a strange quark is involved, the corresponding GOR relation may not be
a reliable constraint.
For kaon, one may start from the kaon PCAC relation
\begin{eqnarray}
\langle 0| {\bar s} \gamma_\mu \gamma_5 u | K^+(p) \rangle = i p_\mu f_K\ .
\end{eqnarray}
By taking the divergence and using the soft-kaon
theorem, one arrives at the kaon GOR relation
\begin{eqnarray}
- (m_q + m_s) [\langle {\bar q} q \rangle + \langle {\bar s} s \rangle] 
= m_K^2 f_K^2\ .
\label{kgor}
\end{eqnarray}
Due to the large strange quark mass $m_s$, one may expect
nonnegligible corrections from 
${\cal O} (m_s^2)$ or higher order, and
this kaon GOR relation, though exact to leading order in quark mass,
may not be a good constraining equation for
$m_s$ and $f_K$.
Indeed, using the conventional QCD 
parameters [for example, see Ref.~\cite{ioffe}.]
\begin{eqnarray}
m_s = 150 ~{\rm MeV} \;; \quad
\langle {\bar s} s \rangle = 0.8 \langle {\bar q} q \rangle\ ,
\end{eqnarray}
Eq.(\ref{kgor}) yields $f_K=116$ MeV, which is much smaller than 
its experimental
value $f^{expt}_K \sim 160$ MeV~\cite{epjc}. Much smaller $m_s$  
from the baryon sum rules~\cite{espriu} leads to much larger
discrepancy.
To reproduce $f^{expt}_K$ from this kaon GOR relation, the 
strange quark mass 
(assuming that $\langle {\bar s} s \rangle$ is fixed as above) needs to be 
around 300 MeV, which is evidently too large for the current mass of 
strange quark.  
Consideration for the $\eta_8$ case 
leads to a similar (even larger) discrepancy from the $\eta_8$
GOR relation.

Then one may naturally attempt to improve the GOR relations especially
when strange quark is involved.  
We look for the improvement that provides a reliable relation between
$f_K$ and $m_s$.  This improvement of course should not affect the 
well-established GOR relation in $u, d$ quark sector, i.e. Eq.(\ref{gor}).
We then look for the $m_s$ in this improvement that reproduces 
$f^{expt}_K$.  For this prediction to be reliable,
this $m_s$ must be tested in the other case such as $\eta_8$.

As the discrepancy mentioned above comes from the large strange quark mass,  
it may be natural to calculate 
higher quark mass corrections to the GOR relations.  
For this purpose, we use 
QCD sum rules~\cite{SVZ,qsr} for the axial-vector correlation
function.  Specifically, 
we wish to construct a sum rule whose phenomenological side
contains only the pseudoscalar meson contributions  while 
the QCD side is proportional to the quark masses.  To leading order in 
quark masses, we want it to reproduce the GOR relation precisely so that
we can systematically study the corrections from higher orders in
quark mass to the GOR relation.
Similar QCD sum rule calculations of the kaon decay constant  
$f_K$~\cite{choe} did not quite look into this higher order 
corrections and our approach in this work is different from
them.

To illustrate our approach briefly,  we start by noting that
the correlation function of the axial-vector current for example 
$A_\mu={\bar u} \gamma_\mu \gamma_5 d$ 
contains the two invariant functions, $\Pi_1$ and $\Pi_2$,
defined by
\begin{eqnarray}
\Pi_{\mu\nu} &=& i\int d^4x e^{ip\cdot x} 
\langle 0| T[ A_\mu (x)~ A_\nu (0) ] |0 \rangle \ ,\nonumber \\ 
&\equiv& -g_{\mu\nu} \Pi_1 + p_\mu p_\nu \Pi_2\ .
\end{eqnarray}
This is in contrast to the vector correlation function 
which contains (under the isospin symmetry) one invariant function
due to the current conservation.
PCAC tells us that pion and its higher resonances contribute
only to $\Pi_2$ while the axial-vector meson contributions are contained
in both $\Pi_1$ and $\Pi_2$.  PCAC further imposes that
$p^\mu \Pi^A_{\mu\nu}$ picks up contributions only from
the pionic resonances. 
Therefore, to investigate the properties of a pion without having
contamination from the $a_1$ meson,
it may be useful to consider the correlation
function defined by
\begin{eqnarray} 
\Pi(p^2) \equiv -{p^\mu p^\nu \over p^2} \Pi_{\mu\nu}
=\Pi_1 - p^2 \Pi_2\ .
\label{ax1}
\end{eqnarray}
This projected correlation function in the QCD side 
must be proportional to quark masses. Therefore, the 
sum rule from $\Pi(p^2)$ may 
provide a useful constraint which directly relates quark masses to
the pion decay constant.  
By generalizing this sum rule to the
strange quark sector, we investigate 
the quark mass corrections of the order ${\cal O}(m_s^2)$ or
higher to the corresponding GOR relation.
This may give stringent constraints for $m_s$.

This paper is organized as follows. In Section~\ref{sec:general},
we present the OPE calculation for the projected correlation function
of the general axial-vector current ${\bar q_1} \gamma_\mu \gamma_5 q_2$.
Using this OPE up to dimension 6, we construct QCD sum rules for $f_\pi$
in  Section~\ref{sec:pion} and estimate how large the higher
order corrections in quark mass to the GOR relation.
In Section~\ref{sec:kaon}, we apply this framework to the kaon
sum rule.  We use this kaon sum rule to determine the strange quark 
mass $m_s$ by using $f_K^{expt}$ as an input.
In Section~\ref{sec:eta}, we test this $m_s$ in the $\eta_8$ sum rule and see if
it reproduces $f_{\eta_8}$  consistent with the phenomenology. 
We summarize in Section~\ref{sec:sum}

\section{The projected sum rule for the general axial-vector current}
\label{sec:general}

In this section, we calculate the operator product expansion (OPE) for 
the projected correlation function
of the general axial-vector current ${\bar q_1} \gamma_\mu \gamma_5 q_2$ 
\begin{eqnarray}
\Pi^A (p^2) &\equiv& -{p_\mu p_\nu \over p^2}~ \Pi^A_{\mu\nu}  \nonumber \\ 
&=& -{p_\mu p_\nu \over p^2}~i \int d^4 x~ e^{ip\cdot x}~ 
\langle 0| 
T[{\bar q_1} (x) \gamma_\mu \gamma_5 q_2(x)~ 
{\bar q_2}(0) \gamma_\nu \gamma_5 q_1 (0) ]
|0 \rangle \nonumber \\
&=& {i\over p^2} \int d^4x e^{ip\cdot x}
{\rm Tr}\left [ 
\fslash{p} \gamma_5 iS_2(x,0) \fslash{p}\gamma_5 iS_1(0,x) \right ]\ , 
\label{cor}
\end{eqnarray}
where $q_j (j=1,2) = u,d~{\rm or}~ s$, and $iS_j(x,0)$ is the quark 
propagator of the flavor $q_j$ in the
fixed-point gauge.  Once the OPE of this
correlation function is calculated, it will be straightforward
to obtain the OPE for pion, kaon and $\eta_8$.
The advantage of using this projected correlation function 
is that the axial-vector meson contributions are canceled 
and the continuum threshold starts at a larger scale so that
the correlation function is well saturated by the low-lying resonance.

A few technical remarks are in order before we proceed to the detailed
OPE calculation.  As $\Pi^A(p^2)$ is a scalar function, the OPE involves only
even dimensional operators. No odd dimensional operators can contribute
to the sum rule.  Also, as the axial-vector current is not conserved
by the finite quark mass, all the OPE must be proportional to 
the quark masses.  Thus we can focus on the operators
that contain quark masses.  This sum rule in the end will give 
sensitive constraints relating
quark mass to the corresponding meson decay constant.

The OPE of the correlator Eq.(\ref{cor})
can be calculated by the standard techniques in the fixed-point
gauge~\cite{fort}.
To include the finite quark mass effectively, we calculate 
the correlator in the momentum space, 
\begin{eqnarray}
\Pi^A (p^2) = {i \over p^2} \int {d^4 k \over (2\pi)^4}  {\rm Tr} \left [
\fslash{p} \gamma_5 iS_1 (k) \fslash{p} \gamma_5 iS_2 (k-p)
\right ] \ .
\label{pcor}
\end{eqnarray}
In the coordinate space, one has to expand the propagator in $m_j$
and hence it is difficult to keep the quark mass terms to all orders
at each dimension. 
The perturbative part can be
calculated straightforwardly using the free quark
propagator
\begin{eqnarray}
iS_j^{free} (k) = i {\fslash {k} + m_{j} \over k^2 - m^2_{j}}~~~(j=1,2)\ .
\label{free}
\end{eqnarray}
Putting this into Eq.~(\ref{pcor}) and 
using the standard techniques of the dimensional regularization and
the Feynman parameterization, we readily compute the perturbative part 
of Eq.~(\ref{pcor}), 
\begin{eqnarray}
\Pi^A_{pert} (p^2) =&&
{3 \over 4\pi^2} \int^1_0 du [m_1^2-u(m_1^2-m_2^2)+m_1 m_2]\nonumber \\
&&\times {\rm ln} [-u(1-u)p^2 +m_1^2 - u (m_1^2-m_2^2) ]\ .
\label{pert}
\end{eqnarray}
Note, this is an order of ${\cal O} (m_j^2)$ or higher.
For the vector correlation function, the $m_1 m_2$ term will 
have the negative sign so that when the two quark flavors are equal 
($q_1=q_2$), the perturbative
part vanishes as can be expected from the current conservation.

For the nonperturbative part, various even dimensional operators can 
contribute to the sum rule.  In this work, we will calculate the OPE up to 
dimension 6.  First we consider the case when one quark propagator in
Eq.~(\ref{pcor}) is
disconnected to form quark condensate~\footnote{In our notation,
$\alpha,\beta,\gamma \cdot \cdot \cdot$ refer Dirac indices 
and $a,b,c \cdot \cdot \cdot$ color indices.}
$iS_{ab}^{\alpha \beta} (k) \rightarrow \int d^4x~ e^{ik\cdot x}  
\langle q^\alpha_a (x) {\bar q}^\beta_b (0) \rangle$ while the
other is remained to be the free propagator.
By performing the Tayler expansion around $x_\mu =0$ and
the Fourier transformation to the momentum space, we obtain
the expansion for the quark condensate~\cite{griegel,elias},
\begin{eqnarray}
\int d^4x&~& e^{ik\cdot x} \langle q^\alpha_a (x) {\bar q}^\beta_b (0) \rangle
= -{\delta^{ab} \over 12} (2\pi)^4 \nonumber \\
&&\times \Bigg \{
\delta^{\alpha \beta} \Big [ \langle {\bar q} q \rangle \delta^{(4)} (k)
+\langle {\bar q} D_\mu q \rangle {\partial \over i\partial k^\mu}
\delta^{(4)} (k) 
+ {1\over 2} \langle {\bar q} D_\mu D_\nu q \rangle 
{\partial \over i\partial k^\mu}{\partial \over i\partial k^\nu}
\delta^{(4)} (k) + \cdot \cdot \cdot \Big ] \nonumber \\
&&+ \gamma_\lambda^{\alpha \beta} \Big [\langle {\bar q}\gamma^\lambda
 q \rangle \delta^{(4)} (k) 
+\langle {\bar q}\gamma^\lambda D_\mu q \rangle 
{\partial \over i\partial p^\mu}
\delta^{(4)} (k) +  \cdot \cdot \cdot \Big ] 
\Bigg \} \ .
\label{qqcond}
\end{eqnarray}
One technical remark~\cite{griegel,elias} is that the Wilson
coefficient of the condensate 
$\langle {\bar q} D_{\mu_1} \cdot \cdot \cdot D_{\mu_n} q \rangle$
is related to the Wilson coefficient of $\langle {\bar q} q \rangle$
via
\begin{eqnarray}
C_{{\bar q} D_{\mu_1} \cdot \cdot \cdot D_{\mu_n} q} (k)
={(-i)^n\over n!}\left [ {\partial \over \partial k_{\mu_1}}
\cdot \cdot \cdot {\partial \over \partial k_{\mu_n}} \right ] 
C_{{\bar q} q} (k) \ ,
\end{eqnarray}
and similarly
\begin{eqnarray}
C_{{\bar q} \gamma_\mu  D_{\mu_1} \cdot \cdot \cdot D_{\mu_n} q} (k)
={(-i)^n\over n!}\left [ {\partial \over \partial k_{\mu_1}}
\cdot \cdot \cdot {\partial \over \partial k_{\mu_n}} \right ]
C_{{\bar q}\gamma_{\mu} q} (k)\ .
\end{eqnarray}
Using this technique, we calculate the OPE involving the
quark condensate up to dimension 6. Recalling that
only even dimensional condensates contribute to the sum rule,
the $\delta^{\alpha \beta}$ part of Eq.~(\ref{qqcond})
gives nonzero contributions when the quark-mass part 
of Eq.~(\ref{free}) is taken for the remaining propagator.
Similarly, the $\gamma_\lambda^{\alpha \beta}$ part of Eq.~(\ref{qqcond}) 
contributes when the $\fslash {k}$ part of the free propagator is taken. 
Also, since the OPE must be proportional to quark mass  
we need to extract only the quark mass dependent part from various quark 
condensates ($j=1,2$),
\begin{eqnarray}
\langle {\bar q_j} \gamma_\lambda D_\alpha q_j \rangle
&\rightarrow& -g_{\lambda \alpha } {i m_j \over 4} 
\langle {\bar q_j}q_j \rangle \ ,
\nonumber \\
\langle {\bar q_j} D_\alpha D_\beta q_j \rangle &\rightarrow& 
-g_{\alpha \beta} {m^2_j \over 4} \langle {\bar q_j}q_j \rangle \ ,
\nonumber \\
\langle {\bar q_j} \gamma_\lambda D_\alpha D_\beta D_\sigma q_j \rangle
&\rightarrow& \left ( g_{\lambda \alpha} g_{\beta\sigma}
+g_{\lambda\beta} g_{\alpha\sigma} + g_{\lambda \sigma} g_{\alpha \beta}
\right )
{i m_j^3\over 24} \langle {\bar q_j}q_j \rangle\ .
\end{eqnarray}

In doing so, we obtain the OPE containing the quark condensate
up to dimension 6,
\begin{eqnarray}
\Pi^A_{{\bar q} q}(p^2)
&=& -(m_1 + m_2) \left [ {\langle {\bar q_1} q_1 \rangle \over p^2 -m_2^2}
+ {\langle {\bar q_2} q_2 \rangle \over p^2 -m_1^2} \right ]
\nonumber \\
&+& {1 \over 2} (m_1^2-m_2^2) \left [
{m_1 \langle {\bar q_1} q_1 \rangle \over (p^2 -m_2^2)^2}
-{m_2 \langle {\bar q_2} q_2 \rangle \over (p^2 -m_1^2)^2} \right ]\ .
\label{qcond}
\end{eqnarray}
The corresponding OPE for the  vector correlation function
is obtained by replacing $m_2 \rightarrow -m_2$ and
$\langle {\bar q_2} q_2 \rangle \rightarrow -
\langle {\bar q_2} q_2 \rangle$.
The first term is the one that leads to 
the GOR relation at the order ${\cal O} (m_j)$.
In our calculation we kept the quark mass in the 
denominator of the free quark propagator.  Thus, when we analytically 
continue to the time-like region, the imaginary part picks up the pole at
$p^2 = m_j^2$.  This aspect is
slightly different from the usual application of QCD sum rules
where the pole of the nonperturbative OPE is located at $p^2=0$.
The modification of a sum rule result due to this is marginal in usual
practices.  However, in our sum rule with $\Pi (p^2)$, we are looking at 
small strength both in the phenomenological and QCD side. In this
case, this small distinction due to the quark mass in the
quark propagator could be important especially
when strange quark is involved. 
 
The contribution from the dimension-4 gluon condensate $\left \langle 
{\alpha_s \over \pi} {\cal G}^2 \right \rangle$ should be vanished 
in our sum rule as it is not proportional to quark mass.
The other possible dimension 6 operators would be 
$m_i m_j \left \langle {\alpha_s \over \pi} {\cal G}^2
\right \rangle$ ($i,j=1,2$) 
and $m_{j} \langle {\bar q_{j}} g_s \sigma \cdot {\cal G} 
q_{j} \rangle$. These however do not contribute to our sum rule.
The disappearance of the $m_{j} \langle {\bar q_{j}} 
g_s \sigma \cdot {\cal G} q_{j} \rangle$ contribution can be easily 
understood by 
recalling that the GOR relation is exact at the order 
${\cal O} (m_j)$.  If $m_{j} \langle {\bar q_{j}} g_s \sigma \cdot {\cal G} 
q_{j} \rangle$ does contribute, then the GOR relation is no longer
valid even at the order ${\cal O} (m_j)$.
Indeed, one can 
show by a direct calculation that 
$m_{j} \langle {\bar q_{j}} g_s \sigma \cdot {\cal G} 
q_{j} \rangle$ does not contribute to our sum rule~\footnote{To check this
easily, it is useful to use the quark propagator in the
coordinate space as given in Eq.(8) and Eq.(13) of
the first reference in Ref.~\cite{wilson}. The two contributions to this 
dimension 6 condensate indeed cancel each other in our sum rule.}.

Verifying the disappearance of the $m_i m_j \left \langle {\alpha_s \over \pi} 
{\cal G}^2 \right \rangle$ contribution (more precisely its imaginary part) 
is technically more involved.
For a consistent calculation, one needs to consider 
the quark propagators with one and two gluons attached. 
Using the quark propagators interacting with gluons (see Ref.~\cite{qsr}
for their explicit expressions),
we readily compute this contribution and obtain
\begin{eqnarray}
&&-{1\over 8} \left \langle {\alpha_s \over \pi} 
{\cal G}^2 \right \rangle
\int^1_0 du \left [ (1-u)^2 m_1 + u^2 m_2 \right ] {m_1+m_2 \over L^2} 
\nonumber \\
&&{\rm where}~~
L = u(1-u) p^2 -m_1^2 + u(m_1^2-m_2^2)\ .
\end{eqnarray}
Note, the numerator is already dimension 6. Thus, up to dimension 6, we
can safely drop the quark mass dependence in $L$. 
Now the imaginary part of the first term involving $(1-u)^2 m_1$ is 
proportional to $\int^1_0~ du~ \delta^\prime (up^2)$ which of course vanishes.
By the same reasoning, the second term involving $u^2 m_2$ also
does not pick up the imaginary part.
Therefore, the dimension 6 contribution involving the gluon condensate does
not contribute to our sum rule.

Before closing this section, here we give the imaginary part of the OPE 
[Eqs.(\ref{pert}) and (\ref{qcond})] 
up to dimension 6 for the correlation function of the
general axial-vector current
$A^\mu ={\bar q}_1 \gamma^\mu \gamma_5 q_2$,
\begin{eqnarray}
&&{1 \over \pi} {\rm Im} \Pi (s) \nonumber \\
&=&
-{3\over 8\pi^2} 
\sqrt{1-{2(m_1^2+m_2^2) \over s}+{(m_1^2-m_2^2)^2 \over s^2}} 
\left [(m_1+m_2)^2 -{(m_1^2-m_2^2)^2 \over s} \right ] 
\theta\left[ s-(m_1+m_2)^2 \right] \nonumber \\
&+&(m_1+m_2) \left [ \langle{\bar q_1} q_1 \rangle \delta(s-m_2^2)
+\langle{\bar q_2} q_2 \rangle \delta(s-m_1^2) \right ]
\nonumber \\
&+& 
{1 \over 2} (m_1^2-m_2^2) \left [
m_1 \langle {\bar q_1} q_1 \rangle {d\over ds} \delta (s -m_2^2)
- m_2 \langle {\bar q_2} q_2 \rangle {d\over ds} \delta (s -m_1^2) \right ]\ .
\label{gope}
\end{eqnarray}
Note, this is symmetric under $1 \leftrightarrow 2$. From this,
it will be straightforward to obtain the OPE for
the pion, kaon and $\eta$ sum rules to be discussed below. 

Now, the advantage of working with the projected correlation function
becomes clear.
This OPE contains a few sources of uncertainties, 
quark condensates and quark masses. Other sources of uncertainty 
such as quark-gluon mixed condensate, gluon condensate,
or four-quark operator do not
appear in our sum rule, which will certainly reduce the error in
our prediction. Furthermore, $\langle {\bar q} q \rangle$ ($q=u,d$) 
can be  determined rather accurately from the pion GOR 
relation Eq.~(\ref{gor}) once
$m_q$ is fixed.  
As we will see in the next section, the higher order corrections 
in quark mass are small in this $u, d$ sector. 
When strange quark is involved, we have only the two 
additional QCD parameters, $m_s$ and $\langle {\bar s} s \rangle$.
Thus, as long as it is stable under the
modeling of higher resonances or Borel mass variation, 
this sum rule can give a stringent constraint for $m_s$.

\section{The pion sum rule}
\label{sec:pion}

Having calculated the OPE for the general axial-vector correlator,
we can easily obtain the pion sum rule. 
To do this, we set $q_1=u$, $q_2=d$ in Eq.(\ref{gope}).
As the normal GOR relation Eq.~(\ref{gor}) is well established, 
our improvement including higher orders in quark masses
is not expected to be large for this pion case. In fact,
the small quark masses can guarantee this expectation but
we will check this for completeness. 

Using the isospin symmetry,
$m_u=m_d\equiv m_q$ and $\langle {\bar u} u \rangle = 
\langle {\bar d} d \rangle \equiv \langle {\bar q} q \rangle$,
we obtain the spectral density (the imaginary part) of the OPE up to 
dimension 6 ,
\begin{eqnarray}
{1 \over \pi} {\rm Im} \Pi^{ope}_{\pi} (s) =
-{3m_q^2\over 2\pi^2} 
\sqrt{1-{4m^2_q \over s}}~~  
\theta ( s-4m_q^2 ) 
+4m_q \langle{\bar q} q \rangle \delta(s-m_q^2)\ .
\end{eqnarray}
The phenomenological spectral density has pion and higher resonances of pion,
namely,
\begin{eqnarray}
{1 \over \pi} {\rm Im} \Pi^{phen}_{\pi} (s)= 
-f_\pi^2 m_\pi^2 \delta(s-m_\pi^2) + (\pi^\prime~{\rm contribution}) +
\cdot \cdot \cdot\ .
\end{eqnarray}
Note that there is no $a_1$ meson contribution in our projected
sum rule.
Matching the two expressions under the Borel transformation and
invoking QCD duality for higher resonances, we obtain the pion sum rule
\begin{eqnarray}
&&\int^{S_0}_0 ds e^{-s/M^2} {1 \over \pi} {\rm Im} \left [ \Pi^{ope}_{\pi} (s)
-\Pi^{phen}_{\pi} (s)\right ] =0 \nonumber \\
&&\rightarrow f_\pi^2 m_\pi^2 e^{-m_\pi^2/M^2} =
{3m_q^2 \over 2\pi^2} \int^{S_0}_0 ds~ e^{-s/M^2} \sqrt{1-{4m^2_q \over s}}~~  
\theta (s-4m_q^2 )
- 4m_q \langle{\bar q} q \rangle e^{-m_q^2/M^2} \ .
\label{psum}
\end{eqnarray}
The first term (the perturbative part) is an order ${\cal O}(m_q^2)$ or
higher.  To the order ${\cal O}(m_q)$, we recover
precisely the GOR relation Eq.(\ref{gor}) as expected. Therefore,
the perturbative part is the correction to the GOR relation.

\begin{figure}
\caption{ 
The Borel curve for the pion decay constant. 
The solid (straight) line is the $f_\pi$ from the GOR relation and
the dashed (curved) line is the $f_\pi$ from the Borel sum rule
of Eq.~(\ref{psum}). 
}
\label{fig1}

\setlength{\textwidth}{6.1in}   
\setlength{\textheight}{9.in}  
\centerline{%
\vbox to 2.4in{\vss
   \hbox to 3.3in{\includegraphics{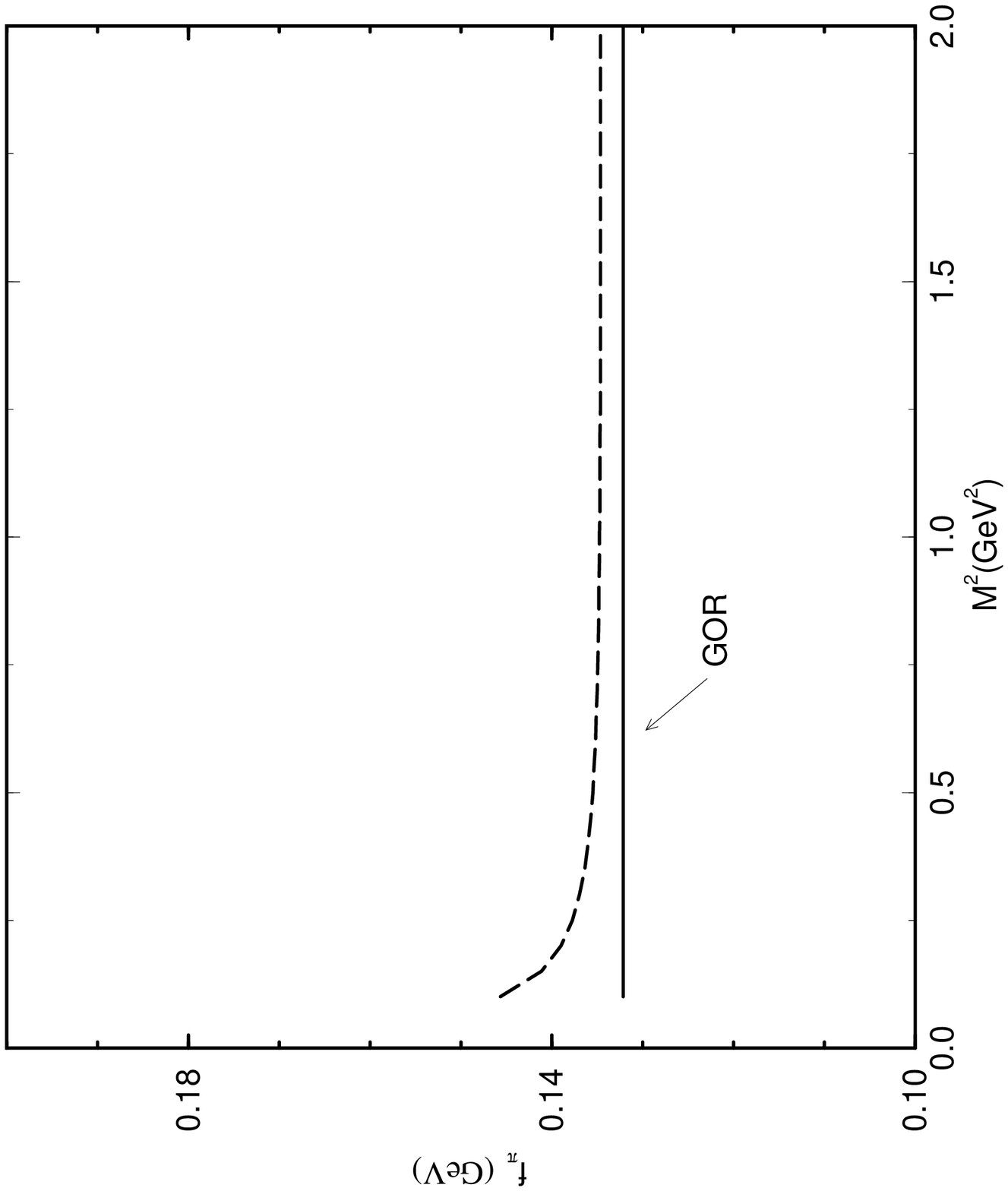}\hss}}
}
\vspace{100pt}
\eject
\end{figure}

We use this sum rule to plot the Borel curve for 
$f_\pi$ setting $m_\pi =138$ MeV, $m_q =7.2$ MeV~\cite{gasser,leut}.
The GOR relation gives (assuming $f_\pi=131$ MeV) 
$\langle{\bar q} q \rangle =-(225~{\rm MeV})^3$. 
The continuum threshold is taken to be $S_0=1.7$  
corresponding to the $\pi^\prime$ mass. 
Figure~\ref{fig1} shows the Borel curve for $f_\pi$, which is
quite stable with respect to the variation of the Borel mass $M^2$.
Furthermore, the resulting $f_\pi$ is not sensitive to $S_0$.
Also shown by the straight line is the $f_\pi$ obtained from the  
GOR relation.  The deviation
from GOR relation is about 3 \%. Thus our sum rule including
the higher mass corrections 
does not change the established pion GOR relation.

\section{The kaon sum rule--to constrain the strange quark mass}
\label{sec:kaon}

\begin{figure}
\caption{ 
The kaon decay constants for given $m_s$ are plotted with
respect to the Borel mass. For each $m_s$, we plot the
curve from the kaon GOR relation and the one from the kaon sum rule. 
The solid lines are for $m_s=100$ MeV,
the dashed lines are for $m_s=150$ MeV and
the dot-dashed lines are for $m_2 =200$ MeV.
The parallel straight lines are obtained from the kaon GOR
relation for given $m_s$ while the corresponding curved
lines are from the kaon Borel sum rule Eq.~(\ref{ksum}). 
The Borel curve (or kaon GOR relation) is shifted noticeably
as we change $m_s$.  Also, higher order corrections of quark
mass give 20 $\sim$ 25 \% shift from the corresponding GOR value. 
The $f_K$ from the Borel curves is closer to its experimental
value of 160 MeV.
}
\label{fig2}

\setlength{\textwidth}{6.1in}   
\setlength{\textheight}{9.in}  
\centerline{%
\vbox to 2.4in{\vss
   \hbox to 3.3in{\includegraphics{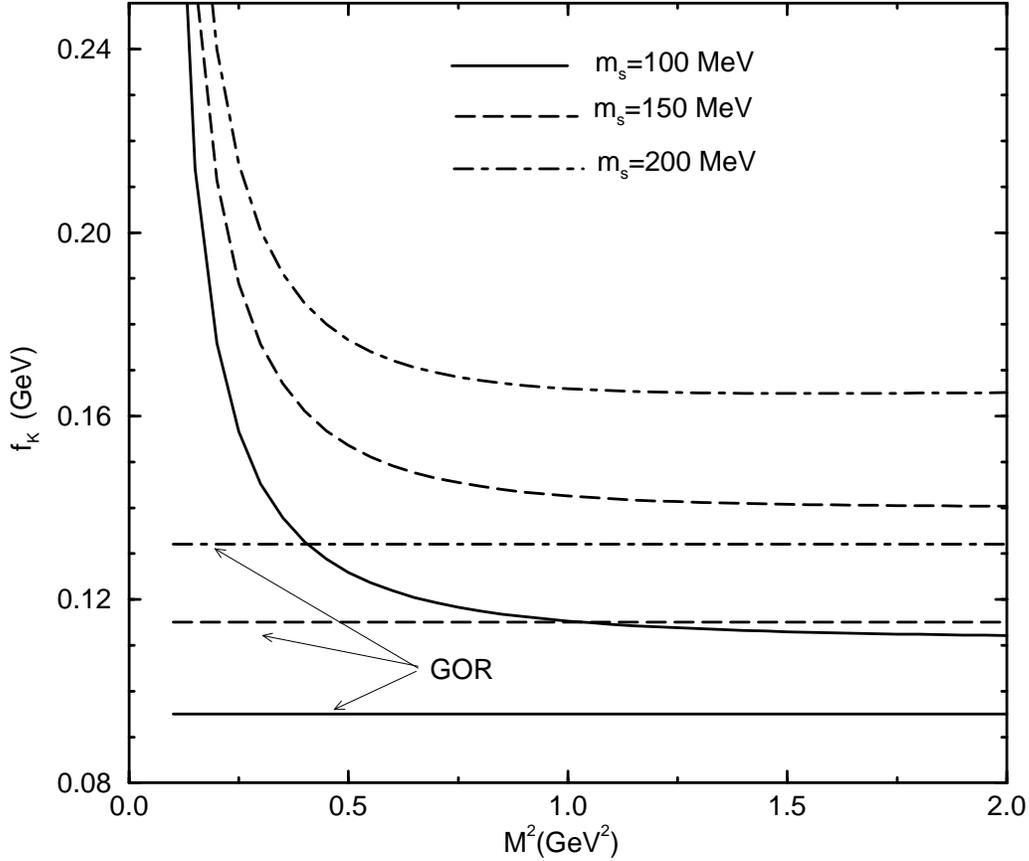}\hss}}
}
\vspace{100pt}
\eject
\end{figure}

We now construct a sum rule for kaon. The small
deviation from the GOR relation observed in the pion sum rule
can not be guaranteed in this case because of the large 
strange quark mass.
In this sum rule, we set $q_1= q$ (here $q=u~{\rm or}~d$)
and $q_2=s$ in Eq.~(\ref{gope}).
The similar procedure as before leads to the kaon sum rule
\begin{eqnarray}
f_K^2 m_K^2 e^{-m_K^2 /M^2}  
&=& {3 \over 8\pi^2} \int^{S_0}_{(m_q+m_s)^2} ds~ e^{-s/M^2}
\sqrt{1-{2(m_q^2+m_s^2) \over s}+{(m_q^2-m_s^2)^2 \over s^2}} 
\nonumber \\
&\times& \left [(m_q+m_s)^2 -{(m_q^2-m_s^2)^2 \over s} \right ] 
\theta\left[ s-(m_q+m_s)^2 \right] \nonumber \\
&-&(m_q+m_s) \left [ \langle{\bar q} q \rangle e^{-m_s^2/M^2} 
+\langle{\bar s} s \rangle e^{-m_q^2/M^2} \right ]
\nonumber \\
&-& 
{1 \over 2} (m_q^2-m_s^2) \left [
m_q \langle {\bar q} q \rangle {1\over M^2} e^{-m_s^2/M^2} 
-m_s \langle {\bar s} s \rangle {1\over M^2} e^{ -m_q^2/M^2} \right ]\ .
\label{ksum}
\end{eqnarray}
Again, to leading order in $m_q$ (or $m_s$),  we recover
the kaon GOR relation Eq.~(\ref{kgor}).  The first and
third terms are therefore the higher order corrections to the kaon GOR relation.
To estimate how large the corrections are, 
we plot for $f_K$ with respect to $M^2$ in
figure~\ref{fig2}. In this plot, we use 
$m_K = 494$ MeV and the continuum threshold is set to $S_0 =2$ GeV$^2$
corresponding to $K_0^*$ (1430)~\footnote{However,
the sensitivity to $S_0$ is very weak. For $S_0=2.5$ GeV$^2$,
the result is changed only by 2 \% .}. For the strange quark condensate,
we set $\langle {\bar s} s \rangle = 0.8\langle {\bar q} q \rangle$~\cite{qsr}.
As for the strange quark mass, a wide range of its value can
be found in literatures~\cite{maltman,dom,colangelo,chet,jamin,narison}.
It is within the range $m_s\sim 100 - 200$ MeV.
To cover this range of $m_s$,
we plot the Borel curves for $m_s= 100$ (solid), 150 (dashed), 200 
(dot-dashed) MeV.  The corresponding straight lines
are the ones from the kaon GOR relation Eq.~(\ref{kgor}).
We see that all three Borel curves become quite flat for $M^2 \ge 1$ GeV$^2$,
but the extracted $f_K$ is quite different from 
the $f_K$ given by the kaon GOR relation, about 20 -25 \% level.
The corrections shift the Borel curves upward and it is clear
that the $f_K$
extracted from them is closer to its experimental value 
$f^{expt}_{K}\sim 160$ MeV. Thus our improvement helps to
achieve a better agreement with the experiment. To reproduce 
$f^{expt}_{K}$ only from the kaon GOR relation, $m_s$ should
be abnormally large like $m_s \sim 300$ MeV, which is obviously  
not acceptable.

Most interesting feature is that the stable Borel curve is 
shifted noticeably as we change $m_s$. 
It means that, once $f_K$ is fixed, this sum rule
can give a stringent constraint for $m_s$ or vice versa.
We therefore look for $m_s$ which reproduces the experimental
kaon decay constant $f^{expt}_K \sim 160$. 
Since $m_q$ and  $\langle{\bar q} q \rangle $ are well fixed by the GOR
relation for pion, only uncertainty in our sum rule
comes from $\langle{\bar s} s \rangle $.  According to the baryon 
sum rules~\cite{ioffe,rry}, the strange quark condensate at 1 GeV is
estimated as
\begin{eqnarray}
\langle{\bar s} s \rangle= (0.8 \pm 0.1)\langle{\bar q} q \rangle \ . 
\end{eqnarray}
Using this input,  we obtain the strange quark mass that reproduces the
experimental $f_K$ at $M^2=1$ GeV$^2$,  
\begin{eqnarray}
m_s = 186.5 \pm 8.5~ {\rm MeV}\ .
\end{eqnarray}
The error comes from the uncertainty in $\langle{\bar s} s \rangle$.
This $m_s$ is the value  at 1 GeV scale because other parameters used in 
extracting this is defined at 1 GeV. The Borel stability in extracting
this is fairly good. 
Even if we take into account the additional error of an order 5 MeV 
from the continuum threshold $S_0$, the extracted $m_s$ from this sum 
rule is sufficiently accurate.  Higher  
dimensional operators starts at dimension 8, which are
expected to be small.
Our $m_s$ is somewhat larger than the conventional value of $m_s =150$ MeV or
the value from hadronic $\tau$ decay $m_s=159$ MeV~\cite{maltman}.
But it is smaller than the one from Ref.~\cite{chet}  and 
is somewhat consistent with the value from the current algebra
ratio $m_s=175 \pm 25$ MeV~\cite{leut} or with the one in Ref.~\cite{jamin}. 

Possible modification to our results can be expected from $\alpha_s$ correction
to the perturbative part. Note however that
the dominant piece to our sum rule is the dimension 4 OPE containing
quark condensates. Contribution from the perturbative part, which is 
an order ${\cal O} (m_q^2)$, is about 20 \% level in our kaon sum rule.
The $\alpha_s$ correction is expected to be proportional to $\alpha_s/\pi$
times the perturbative part [For example, see the $\rho$ meson
sum rule in Ref.~\cite{SVZ}.].  At $M=1$ GeV, $\alpha_s$ is 
around 0.3~\cite{SVZ}, which gives the
correction of an order 0.1 to the perturbative part.  Thus, unless it 
involves a large numerical coefficient,
the $\alpha_s$ correction to our results is expected to be small.
Nevertheless, in future,  it will be interesting to calculate 
the $\alpha_s$ correction explicitly and make sure this expectation.

\section{The $\eta_8$ sum rule}
\label{sec:eta}

Having determined $m_s$ from the kaon sum rule, we now test 
whether it gives a consistent result for the $\eta_8$ case
by constructing a sum rule for the current
\begin{eqnarray}
A_\mu^8 = {1\over \sqrt{6}}
\left ( {\bar u} \gamma_\mu \gamma_5 u + {\bar d} \gamma_\mu \gamma_5 d
- 2{\bar s} \gamma_\mu \gamma_5 s \right ) \ .
\end{eqnarray}
We recall that the $\eta_8$ decay constant is defined by the matrix element
\begin{eqnarray}
\langle 0 | A_\mu^8 |\eta_8 (p) \rangle
=if_{\eta_8} p_\mu \ .
\end{eqnarray}
Refs.~\cite{epjc,feldmann} give $f_{\eta_8} \sim 157 -165$ MeV.
Using $m_s$ from the kaon sum rule, we will see if the $f_{\eta_8}$
from our sum rule is consistent within this range.

Taking the divergence of the axial-vector current
and the subsequent use of the soft-$\eta_8$ theorem lead to the following
$\eta_8$ GOR relation
\begin{eqnarray}
f^2_{\eta_8} m^2_{\eta_8} = 
-{4 \over 3} m_q \langle {\bar q} q \rangle 
-{8\over 3} m_s \langle {\bar s} s \rangle\ .
\label{egor}
\end{eqnarray}
Here again, we have used the isospin symmetry $m_u=m_d\equiv m_q$
and $\langle {\bar u} u \rangle = \langle {\bar d} d \rangle
\equiv \langle {\bar q} q \rangle$.
The OPE for the correlation function 
\begin{eqnarray}
\Pi (p^2)=-{p_\mu p_\nu \over p^2}
i \int d^4 x e^{ip\cdot x} \langle 0 | T A^8_\mu (x) A^8_\nu (0) | 0 \rangle
\end{eqnarray}
can be easily obtained from the general formula Eq.~(\ref{gope}).
Assuming that the continuum contribution starts at the $\eta^\prime$
resonance (i.e. $S_0=0.92$ GeV$^2$), we easily obtain the $\eta_8$ sum rule,
\begin{eqnarray}
f^2_{\eta_8} m^2_{\eta_8} e^{-m^2_{\eta_8}/M^2}
&=& {m_q^2 \over 2\pi^2} \int^{S_0}_{4m_q^2} ds~ e^{-s/M^2} 
\sqrt{1-{4m^2_q \over s}}  
- {4 \over 3} m_q \langle{\bar q} q \rangle e^{-m_q^2/M^2}
\nonumber \\
&+& {m_s^2 \over \pi^2} \int^{S_0}_{4m_s^2} ds~ e^{-s/M^2} 
\sqrt{1-{4m^2_s \over s}}~~  
- {8 \over 3} m_s \langle{\bar s} s \rangle e^{-m_s^2/M^2} \ .
\label{esum}
\end{eqnarray}
Again, to leading order in quark mass, this reproduces
the $\eta_8$ GOR relation Eq.~(\ref{egor}).
The perturbative terms that are the second order in quark mass
constitute the higher order corrections.
  
\begin{figure}
\caption{ 
The $\eta_8$ decay constant with respect to to the Borel mass.
In this plot, we use the best fitting $m_s$ from the kaon sum rule.
The solid straight line is obtained from the $\eta_8$ GOR
relation and the thick dashed curve is from the $\eta_8$ sum rule (with
$S_0=0.92$ GeV$^2$)
Eq.~(\ref{esum}).  There is 30 \% change from the $\eta_8$
GOR relation.
The thin dashed line is obtained with $S_0=1.4$ GeV$^2$ to show the 
sensitivity to $S_0$.   
}
\label{fig3}

\setlength{\textwidth}{6.1in}   
\setlength{\textheight}{9.in}  
\centerline{%
\vbox to 2.4in{\vss
   \hbox to 3.3in{\includegraphics{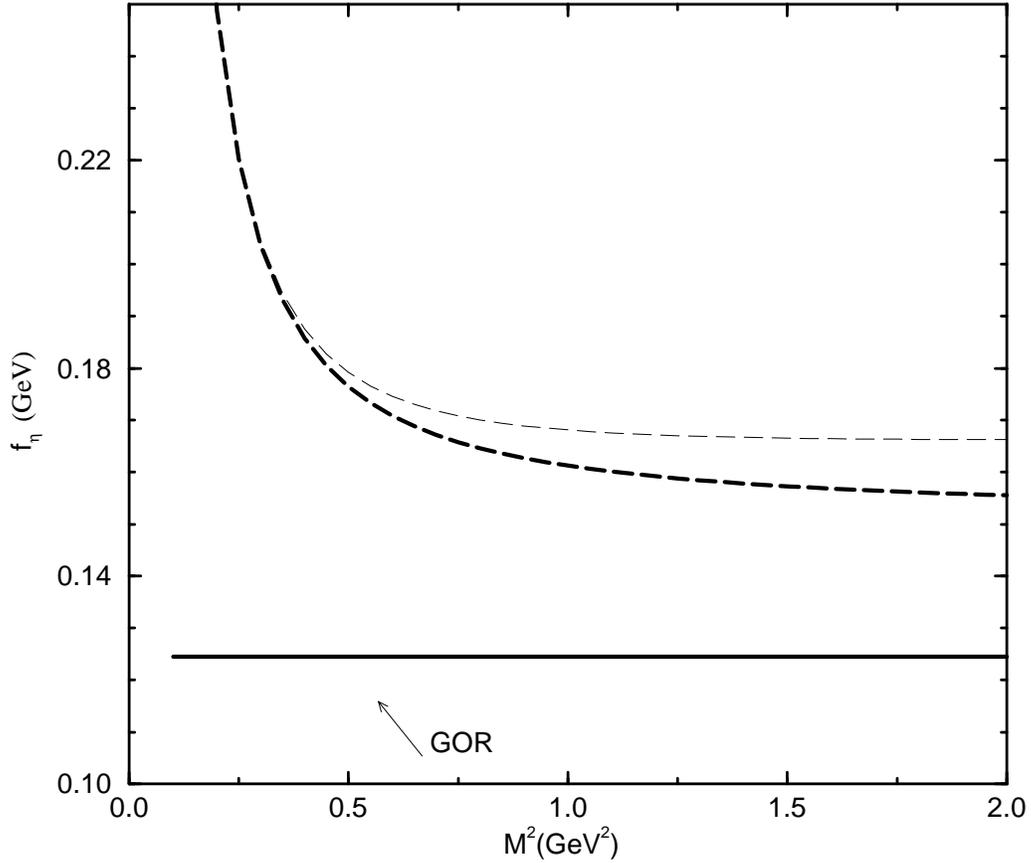}\hss}}
}
\vspace{100pt}
\eject
\end{figure}

The Borel curve for $\eta_8$ is shown by the thick dashed 
line in Figure~\ref{fig3}.  In this plot, we
use $m_s=186$ MeV obtained from the kaon sum rule and
assume that $m_{\eta_8}= m_\eta=547$ MeV.
To show the sensitivity to the
continuum threshold, the curve for $S_0=1.4$ GeV$^2$
is also shown by the thin dashed line.  Around $M^2=1$ GeV$^2$,
the two curves differ by 5\%. 
The Borel stability is not as good as the kaon case 
but $f_{\eta_8}$ becomes only 5 \% smaller at $M^2=2$ GeV$^2$ 
than the one at $M^2=1$ GeV$^2$. 
As can be seen,  the $f_{\eta_8}$ from this sum rule becomes
30 \% larger than what the $\eta_8$ GOR relation gives.
Therefore, $f_{\eta_8}$ picks up a large correction
from higher orders in quark mass  
to the $\eta_8$ GOR relation.  From this sum rule, we extract 
$f_{\eta_8} = 160 $ MeV at $M^2=1$ GeV$^2$.  
This is very close to its value in literatures
$157 - 165 $ MeV ~\cite{epjc,feldmann} even if we
take into account the uncertainties due to the continuum threshold 
or the slight dependence on the Borel mass. 
Therefore, $m_s=186$ MeV extracted from the kaon sum rule consistently
reproduces $f_K$ and $f_{\eta_8}$.

\section{Summary}
\label{sec:sum}

In this work, we have constructed QCD sum rules exclusively for 
decay constants of the 
pseudoscalar mesons, $\pi$, kaon and $\eta_8$. 
The phenomenological side of our sum rule
contains only the pseudoscalar mesons while the QCD side is
proportional to quark masses.  Our sum rules give a very stable 
constraint relating the meson decay constants and the quark masses.
To leading order in quark masses, our sum rules produce the GOR 
relations, which allows us to study the higher order
corrections in quark masses.  From the kaon sum rule, we 
have obtained a sensitive constraint relation 
for $f_K$ and $m_s$, which restricts $m_s \sim 186$ MeV
in order to reproduce the experimental $f_K$.  
Such a $m_s$ when applied to the $\eta_8$ sum rule gives $f_{\eta_8}$
consistent with its value in literatures.

\acknowledgments
This work is supported by the Brain Korea 21 project.
We thank Prof. S. H. Lee for useful discussions.

\end{document}